\documentclass[preprint]{aastex}
\input psfig.sty
\newcommand{\kms}          {\mbox{${\rm km~s^{-1}}$}}
\newcommand{\cc}           {\mbox{${\rm cm^{-3}}$}}

\newcommand{\e}            {\mbox{$^{-1}$}}

\newcommand{\eee}          {\mbox{$^{-3}$}}
\newcommand{\simgt}        {\gtrsim}
\newcommand{\simlt}        {\lesssim}

\newcommand{\dtbr}         {\mbox{$\Delta T_{\rm b-r}$}}
\def\cm2{\mbox{${\rm cm^{-2}}$}}
\def\h2{\mbox{${\rm H}_2$}}
\def\nh2{\mbox{$n_{\rm H_2}$}}
\def\Nh2{\mbox{$N_{{\rm H}_2}$}}
\def\Mh2{\mbox{$M_{{\rm H}_2}$}}
\def\n2hp{\mbox{N$_2$H$^+$}}
\def\c34s{\mbox{C$^{34}$S}}
\def\fs{\hbox{$.\!\!^{\rm s}$}}

\slugcomment{Accepted by the Astrophysical Journal,
February 14$^{\rm th}$, 1999}
\shorttitle{Pressure Flows and Turbulent Dissipation}
\shortauthors{Williams \& Myers}

\begin{document}

\title{EVIDENCE FOR PRESSURE DRIVEN FLOWS AND TURBULENT DISSIPATION IN THE
SERPENS NW CLUSTER}
\author{Jonathan P. Williams\altaffilmark{1}\altaffilmark{2}}
\altaffiltext{1}{National Radio Astronomy Observatory, 949 N. Cherry Ave.,
Tucson AZ 85721}
\altaffiltext{2}{Department of Astronomy, 211 Bryant Space Science Center,
University of Florida, Gainesville, FL 32611; williams@astro.ufl.edu}
\and
\author{Philip C. Myers\altaffilmark{3}}
\altaffiltext{3}{Harvard--Smithsonian Center for Astrophysics,
60 Garden Street, Cambridge, MA 02138; pmyers@cfa.harvard.edu}

\begin{abstract}
\rightskip = 0pt
We have mapped the dense gas distribution and dynamics in the NW region
of the Serpens molecular cloud in the CS(2--1) and \n2hp(1--0) lines and
3~mm continuum using the FCRAO telescope and BIMA interferometer.
7 continuum sources are found.
The \n2hp\ spectra are optically thin and fits to the 7 hyperfine
components are used to determine the distribution of velocity dispersion.
8 cores, 2 with continuum sources, 6 without, lie at a local linewidth
minimum and optical depth maximum.
The CS spectra are optically thick and generally self-absorbed
over the full 0.2~pc extent of the map.
We use the line wings to trace outflows around at least 3,
and possibly 4, of the continuum sources, and the asymmetry in
the self-absorption as a diagnostic of relative
motions between core centers and envelopes.
The quiescent regions with low \n2hp\ linewidth tend to have more
asymmetric CS spectra than the spectra around the continuum sources
indicating higher infall speeds. These regions have typical sizes
$\sim 5000$~AU, linewidths $\sim 0.5$~\kms,
and infall speeds $\sim 0.05$~\kms. The correlation of CS asymmetry with
\n2hp\ velocity dispersion suggests that the inward flows of material
that build up pre-protostellar cores are driven at least partly by a
pressure gradient rather than by gravity alone.
We discuss a scenario for core formation and eventual star forming
collapse through the dissipation of turbulence.
\end{abstract}


%\clearpage
\rightskip = 0pt
\section{Introduction}
Stars form through the collapse of dense cores within molecular clouds.
The detection and measurement of the motions associated
with such star forming collapse appears to be secure
(see reviews by Evans 2000; Myers, Evans, \& Ohashi 2000).
The focus has primarily been on isolated, individual
star forming regions since these are the least complex cases to
understand both observationally and theoretically.
However, the majority of stars form in clusters
(Zinnecker, McCaughrean, \& Wilking 1993) so a broader
understanding of star formation, including such fundamental
issues as the origin of the IMF and the formation of massive stars,
requires the study of how stars form in groups.

Relative to an individual star,
the deeper potential well of a stellar cluster should imply faster,
and possibly easier to measure, inward motions, but observations are
complicated by the generally greater distance to clusters than to individual
star forming regions studied so far, and by the overpowering luminosity
of massive stars that limit the ability to image
lower mass neighbors and thus to obtain a complete view
of cluster birth. For example, observations of the massive cluster
forming region, W49N, have produced evidence for a global
collapse of material onto the cluster as a whole (Welch et al. 1987)
and, at higher resolution, for infall onto bright individual protostars
(Zhang \& Ho 1997). However, it has not been possible to explore
the formation of more moderate mass stars in these objects
because of dynamic range limitations.

In this paper, we present millimeter line and continuum observations
of a cluster forming region in the Serpens
molecular cloud. This region is well suited for an exploration into
the formation processes of stellar clusters because it is nearby
(d=310~pc; de Lara, Chavarria-K, \& Lopez-Molina 1991) and contains
a moderately dense embedded cluster
(stellar density $\simgt 450$~pc\eee; Eiroa \& Casali 1992)
with many highly embedded millimeter wavelength continuum sources
(Testi \& Sargent 1998) but no O or B stars.
%This young embedded cluster has been well studied from optical
%to millimeter wavelengths (see, for example, recent papers by
%Davis et al. 1999; Kaas 1999; McMullin et al. 1999).
In addition, Williams \& Myers (1997) and Mardones (1998)
have found signatures of widespread infall motions in this region.

The combination of proximity and low mass makes it possible to
identify individual star forming condensations and examine the
structure and dynamics of the cluster on a core by core basis.
With this goal, we mapped the cluster in the optically thick
CS(2--1) and thin \n2hp(1--0) lines with the FCRAO and BIMA telescopes.
Their different optical depths allow us to probe cloud velocities
from an outer envelope to the center (Leung \& Brown 1977).
The two species are also formed by different chemical pathways with
abundances that depend on environment and time (Bergin et al. 1997)
and thus their relative intensities offer information on the
physical conditions and chemical age of the cores.
In an earlier paper (Williams \& Myers 1999a) we reported the discovery
of a starless core that appears to be collapsing based on an analysis
of a small part of the maps. Here, we report on the full dataset
and investigate the dynamics of the dense gas across the whole cluster.
We find a number of new cores, some starless, others with continuum
sources, suggesting that new stars are continually being added to
the group. There are spectroscopic signatures of outflow, infall,
and localized dissipation of turbulence throughout the cloud
which offer important clues about the physical processes involved
in cluster formation.
The observations are outlined in \S2 and the data displayed in \S3.
An analysis of the data follows in \S4, and we discuss our findings
in \S5, concluding in \S6.

\section{Observations}
\label{sec:obs}
Singledish maps of \n2hp(1--0) (93.1762650~GHz, F$_1$F$=01-12$)
and CS(2--1) (97.980968~GHz) were made at the
Five College Radio Astronomy Observatory\footnotemark
\footnotetext{FCRAO is supported in part by the
National Science Foundation under grant AST9420159 and is operated
with permission of the Metropolitan District Commission, Commonwealth
of Massachusetts} (FCRAO) 14~m telescope in December 1996
using the QUARRY 15 beam array receiver and the FAAS backend
consisting of 15 autocorrelation spectrometers with
1024 channels set to an effective resolution of 24~kHz (0.06~km/s).
The observations were taken in frequency switching mode and, after
folding, 3rd order baselines were subtracted. The pointing and focus
were checked every 3 hours on nearby SiO maser sources.
The FWHM of the telescope beam is $50''$, and a map covering
$6'\times 8'$ was made at Nyquist ($25''$) spacing.
QUARRY was replaced with the SEQUOIA array in late 1997.
This 16 element array is built with low noise MMIC based amplifiers
and has much improved sensitivity. This enabled us to map the weak
\c34s(2--1) (96.412982~GHz)
line using the same backends and observing technique in March 1998.

Observations were made with the 10
antenna Berkeley-Illinois-Maryland array\footnotemark
\footnotetext{Operated by the University of California at Berkeley,
the University of Illinois, and the University of Maryland,
with support from the National Science Foundation}
(BIMA) for two 8 hour tracks in each line during
April 1997 (CS; C array)
and October/November 1997 (\n2hp; B and C array).
A two field mosaic was made with phase center,
$\alpha(2000)=18^{\rm h}29^{\rm m}47\fs 5,
~\delta(2000)=01^\circ 16'51\farcs 4$, centered close to S68N,
and a second slightly overlapping pointing at
$\Delta\alpha=33\farcs 0, \Delta\delta=-91\farcs 0$,
centered close to SMM1.
Amplitude and phase were calibrated using 4 minute observations of
1751+096 (4.4~Jy) interleaved with each 22 minute integration on source.
The correlator was configured with two sets of 256 channels at
a bandwidth of 12.5~MHz (0.16~\kms\ per channel) in each sideband
and a total continuum bandwidth of 800~MHz.

The data were calibrated and maps produced using standard procedures in
the MIRIAD package. The two pointings were calibrated together but
inverted from the $uv$ plane individually to avoid aliasing since
their centers are separated by more than one primary beam FWHM.
The FCRAO data (after first scaling to common flux units using a
gain of 43.7~Jy~K\e) were then combined
with the BIMA data using maximum entropy deconvolution. The final maps
were produced by linearly mosaicking the two pointings in the $xy$ plane.
The combination of the single dish and interferometer data
results in maps that are fully sampled from the map size
down to the resolution limit (i.e., there is no ``missing flux'').

In addition, we obtained a map of continuum emission by summing
over the line-free channels. This map showed a number of point
sources corresponding to warm dusty envelopes around embedded
protostars. A similar map was obtained, over the entire Serpens
complex, using the OVRO interferometer by Testi \& Sargent (1998).
Our map was of lower sensitivity and in order to make a better
comparison, we were awarded additional time from
February to April 1999 to map the continuum at 110~GHz in a mosaic
consisting of 3 fields overlapping at their primary beam FWHM
over the same region.

The resolution of the final (naturally weighted) maps was
$8\farcs 1\times 5\farcs 6$ at p.a. $+2^\circ$ for the continuum,
$10\farcs 0\times 7\farcs 8$ at p.a. $-72^\circ$ for CS,
and $8\farcs 5\times 4\farcs 6$ at p.a. $+2^\circ$ for \n2hp.
Additional spectral line datasets were created by restoring to a
common $10''$ (3100~AU) beam for analysis and comparison.
The velocity resolution of these maps was 0.16~\kms.

\section{Analysis}
\label{sec:analysis}
\subsection{Continuum and integrated line maps}
\label{subsec:continuum}
\begin{figure}[htpb]
\vskip 0.0in
\centerline{\psfig{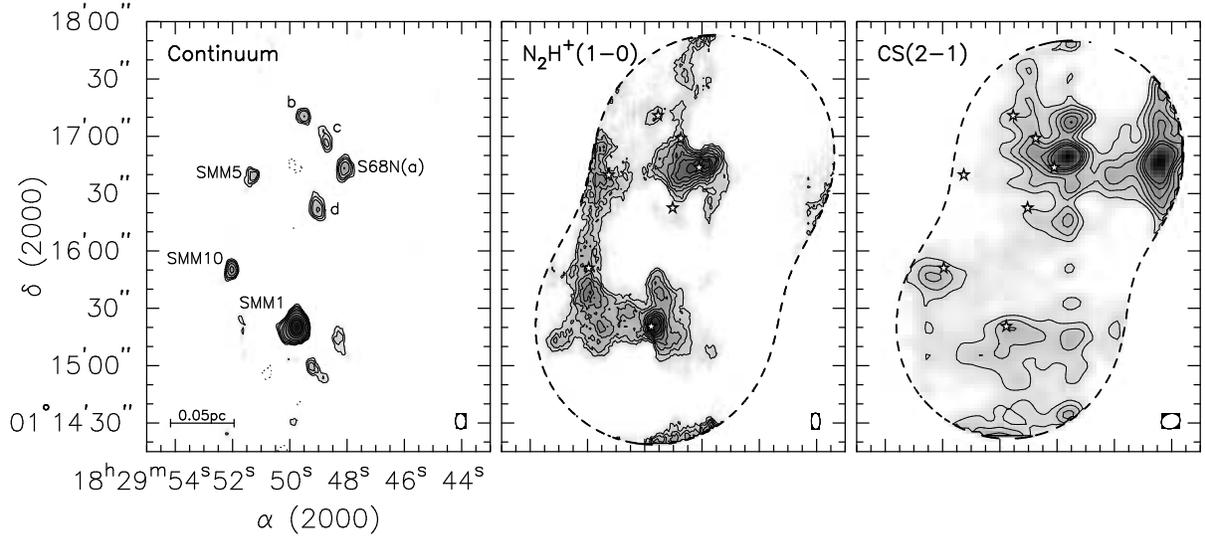}}
\vskip 0.0in
\figcaption[f1.ps]{Continuum and line emission in the Serpens cluster.
The left panel shows the continuum emission at 110~GHz in
a logarithmic greyscale ranging from 5 to 200~mJy~beam\e,
and contours at 7, 9, 11, 15, 20, 30, 50~mJy~beam\e\
(dotted contours are their negative counterparts).
The seven labeled sources (other objects are believed to
be artifacts; see text) are indicated in the two other
panels by the star-shaped symbols.
The center panel shows the combined BIMA+FCRAO \n2hp(1-0) emission
integrated over all hyperfine components from $v=6.5$ to 10.0~\kms.
The greyscale ranges linearly from 2.7 to 36~K~\kms, with contours
at 8.2, 10.9, 13.6, ...K~\kms.
The right panel shows the combined BIMA+FCRAO CS(2--1) emission
integrated from $v=0.75$ to $10.5$~\kms. The greyscale ranges
linearly from 5 to 40~K~\kms, with contours at 10, 12.5, 15, ...K~\kms.
Both line maps have been corrected for primary beam attenuation
and masked beyond the dotted boundaries corresponding to the
50\% level of the two pointings. Synthesized beam sizes are
indicated in the lower right corner for each map.
\label{fig:bima}}
\end{figure}

Maps of the continuum emission and integrated CS and \n2hp\ line
intensities are presented in Figure~\ref{fig:bima}.
The continuum map was obtained
from the 3 field mosaic at 110~GHz (see above). The rms noise level
was 1.0~mJy~beam\e\ at the center, increasing toward the map edges
(all maps are corrected for primary beam attenuation). Seven sources
are labeled; S68N, SMM1 (also known as S68 FIRS1), SMM5, and SMM10
were mapped by Casali, Eiroa, \& Duncan (1993) and Davis et al. (1999)
and we have labeled the three others S68Nb through S68Nd
because of their proximity to S68N.
These seven are also present in Testi \& Sargent (1998)
although we do not confirm several other sources in their map
for which our data have sufficient coverage, sensitivity,
and resolution to detect.
Singledish observations with the 1.3~mm facility bolometer
on the IRAM 30~m telescope should clarify the issue.

Source positions and fluxes are listed in Table~1.
S68Nc may be a double or multiple source since it is highly elongated
but we could not distinguish more than one significant peak,
so it is listed in Table~1 as a single source.
SMM1, with a peak flux of 165~mJy~beam\e,
is considerably brighter than the other sources,
and it proved problematic to clean the map completely of its sidelobes.
Wright et al. (1999) pointed out that systematic errors including
incomplete $uv$ coverage, calibration uncertainties, and pointing errors,
limit the fidelity of a mosaicked image to the true source brightness
distribution to $1-2$\% at best.
Such errors would lie above the level of the noise and therefore may
be misinterpreted as detections.
Both clean and maximum entropy deconvolution methods
(with varying parameters, and source modeling and replacement)
produced maps that all contained the seven labeled sources
but also created elongated features in the vicinity of SMM1
that varied in position and flux from map to map.
Therefore we have a high confidence in the labeled sources
in Figure~\ref{fig:bima}
but we believe the unlabeled features to the southwest of SMM1
to be artifacts of the data acquisition and reduction process
and we disregard them.
The problems associated with this map illustrate the difficulties
in obtaining a complete view of cluster formation in more massive
and luminous star forming environments.

There are also some image artifacts present in the line maps.
The near-zero declination of the source resulted in strong
north-south sidelobes even with the 45 baselines of the BIMA
interferometer. As with the continuum map, these proved difficult
to clean away completely and thus there may be some small sidelobe
contamination from the brighter sources in each map.
Based on experimenting with different deconvolution procedures,
we estimate this uncertainty in line strengths to be $20\%$
in addition to the thermal noise and flux calibration uncertainties.

The \n2hp\ and CS maps each show several condensations but
present a very different appearance. The \n2hp\ map follows
the distribution of the continuum sources much more closely
than the CS map which features a prominent starless core to
the west of S68N (Williams \& Myers 1999a). These differences
between the \n2hp\ and CS maps are probably due to a combination of
differences in optical depth, protostellar outflows, and depletion.
As we show below, the \n2hp\ emission is optically thin but
most CS spectra are self-absorbed and have considerable optical
depth. Moreover, outflow wings are prominent in the CS line
profiles around SMM1 and S68N, but are almost entirely absent
from the \n2hp\ spectra at the same positions.
Finally, the time dependent chemical models of Bergin \& Langer (1997)
suggest that CS should deplete onto
grains prior to star forming collapse but \n2hp\ should remain in
the gas phase even at high densities during the collapse phase.
For these reasons, in the following subsections we determine the
physical properties of the cluster forming gas and individual
star forming cores from the \n2hp\ data and measure
motions between core envelopes and centers using the CS data.

\subsection{\n2hp(1--0)}
\label{subsec:n2hp}
The hyperfine structure of the $J=1-0$ transition of \n2hp\ spreads out
the emission into seven components (Caselli, Myers, \& Thaddeus 1995),
each with considerably lower intensity
than a single component line would have, but with the benefit
of also spreading out the optical depth so that individual
components can be optically thin even when the sum of optical
depths over all components might exceed unity.
By fitting the seven hyperfine components simultaneously,
we maximize the information in the data while taking advantage
of the individual low optical depths to determine the systemic
velocity and linewidth of the gas.

We fit the spectra using a function of the form,
$$T_{\rm B}(v)=\Bigl[J(T_{\rm ex})-J(T_{\rm bg})\Bigr]
               \Bigl\{1-{\rm exp}[-\sum_i g_i\tau(v;v_i)]\Bigr\},
\eqno(1)$$
where the measured brightness temperature $T_{\rm B}$ is a
function of velocity $v$,
$T_{\rm ex}$ is the excitation temperature,
$T_{\rm bg}=2.73$~K is the cosmic background temperature,
and the sum is over hyperfine components $i=1,2,..7$.
For each component, $g_i$ is the statistical weight
(Womack, Ziurys, \& Wyckoff 1992) normalized so $\sum_i g_i=1$,
and $\tau(v;v_i)$ is the total optical depth parameterized by $v_i$,
the relative centroid velocity of each component (Caselli et al. 1995),
$$\tau(v;v_i)=\tau_0\,{\rm exp}[-(v-v_i-v_0)^2/2\sigma^2].
\eqno(2)$$
Here $\tau_0$ is the peak optical depth (summed over all components),
$v_0$ is the systemic velocity of the gas, and $\sigma$ is the
velocity dispersion.

Spectra were analyzed across the map after first restoring to a
circular $10''$ FWHM beam and sampling on a regular $10''$ square grid.
The fits to the spectra require four parameters,
$T_{\rm ex}, \tau_0, v_0$, and $\sigma$, but only the velocity
and dispersion were tightly constrained by the data.
The fitted excitation temperature and optical depth are determined
from the intensities of the 7 hyperfine components but the two are inversely
correlated resulting in a wide range of pairs that fit any given peak with
only small changes in line shape that are indistinguishable given the
moderate signal-to-noise ratios in the data. Moreover, there appear to
be significant excitation anomalies analogous to the very low noise
spectrum in Caselli et al. (1995). Because of the large uncertainties
in $T_{\rm ex}$ and $\tau$, we do not discuss them further.

We also made three component, optically thin, gaussian fits,
$T_{\rm B}(v)=T_0\sum_i g_i\, {\rm exp}[-(v-v_i-v_0)^2/2\sigma^2]$,
to the data. In most cases, the residuals were not significantly greater
than the four parameter fit demonstrating the degeneracy in
$T_{\rm ex}$ and $\tau$. We also checked our fits with the hyperfine
structure fitting routine in the CLASS data reduction package. All 3
methods show good agreement in the velocity and dispersion.

By adding synthetic noise to very low noise \n2hp\ spectra
we tested the effect of varying signal-to-noise ratios on
the fits. The systemic velocity could be accurately determined
even in very noisy spectra but the fitted linewidth systematically
increased as the signal-to-noise ratio decreased. The effect was
noticeable for peak ratios less than 10, and became severe for ratios
less than 5. In the following analysis, only spectra with peak
signal-to-noise ratios greater than 5 are fit. For these spectra,
we derived $v_0$, and $\sigma$ using the method described
by equations (1) and (2). 

\begin{figure}[htpb]
\vskip 0.0in
\centerline{\psfig{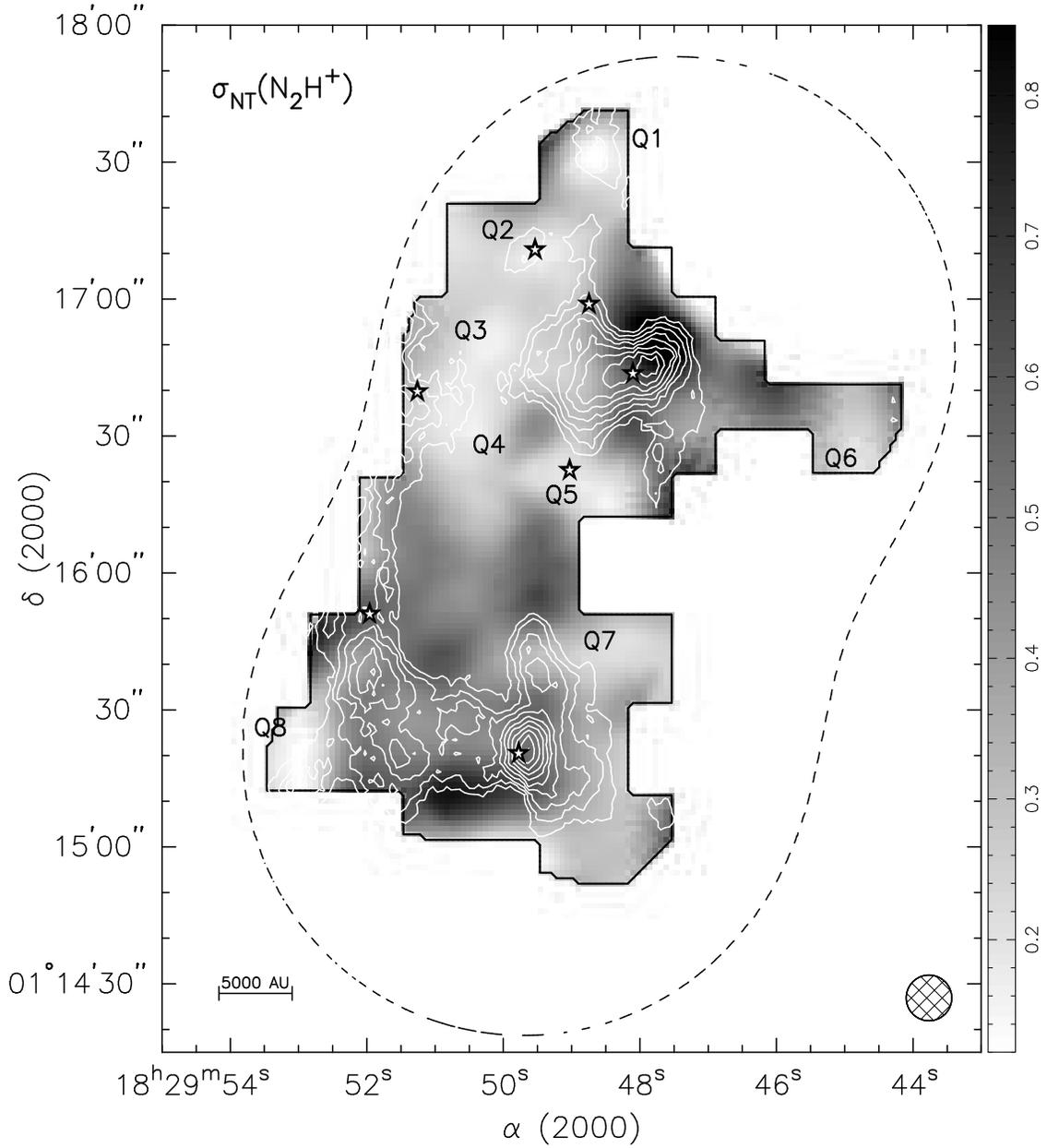}}
\vskip 0.0in
\figcaption[f2.ps]{The non-thermal velocity dispersion of \n2hp\ displayed
in greyscale overlayed on a contour map of integrated
emission as in Figure~1. The dispersion was only determined for
those spectra with a peak signal-to-noise ratio greater than 5 and lie
within the boundary shown by the heavy solid line. The dashed line indicates
the FWHM of the mosaic. The resolution of this map is $10''$ indicated
in the lower right corner. We identify eight regions where the velocity
dispersion is a local minimum, labeled Q1 through Q8.
\label{fig:sig_nt}}
\end{figure}

The systemic velocity varies from 7.9 to 9.3~\kms\
and away from the outflow around S68N (see \S\ref{subsec:cs})
there are no strong gradients.
The dispersion displayed a more interesting variation.
Figure~\ref{fig:sig_nt} plots the non-thermal velocity dispersion,
$\sigma_{\rm NT}=[\sigma^2-\sigma_{\rm T}^2]^{1/2}$,
where $\sigma_{\rm T}=0.075$~\kms\ is the thermal velocity
dispersion for \n2hp\ at $T_{\rm kin}=20$~K (Wolf-Chase et al. 1998).
The greatest values, $\sigma_{\rm NT}>0.6$~\kms\ occur in the cores
containing the S68N and SMM1 sources which both power strong outflows.
The minimum $\sigma_{\rm NT}$ is 0.16~\kms\ which is more than
twice $\sigma_{\rm T}$ and shows that internal motions
in the cores are predominantly turbulent.
However, there are a number of regions where the turbulent
velocity field drops to a local, confined, minimum.
We identify eight such regions and label them Q1 through Q8 (for quiescent).

Three of the quiescent regions, Q1, Q2, and Q8, are coincident with
peaks of integrated \n2hp\ intensity but the other five are not.
Note that even in the low intensity regions, the signal-to-noise
ratio is strong enough to determine the dispersion quite accurately
and that any bias would tend to {\it increase} the dispersion.
The quiescent regions are not prominent in the map of integrated
intensity because their linewidth is small.
Indeed, our analysis of the data suggests to us that these
quiescent ``cores'' are of greater interest than the cores of high
integrated intensity.

\begin{figure}[htpb]
\vskip 0.0in
\centerline{\psfig{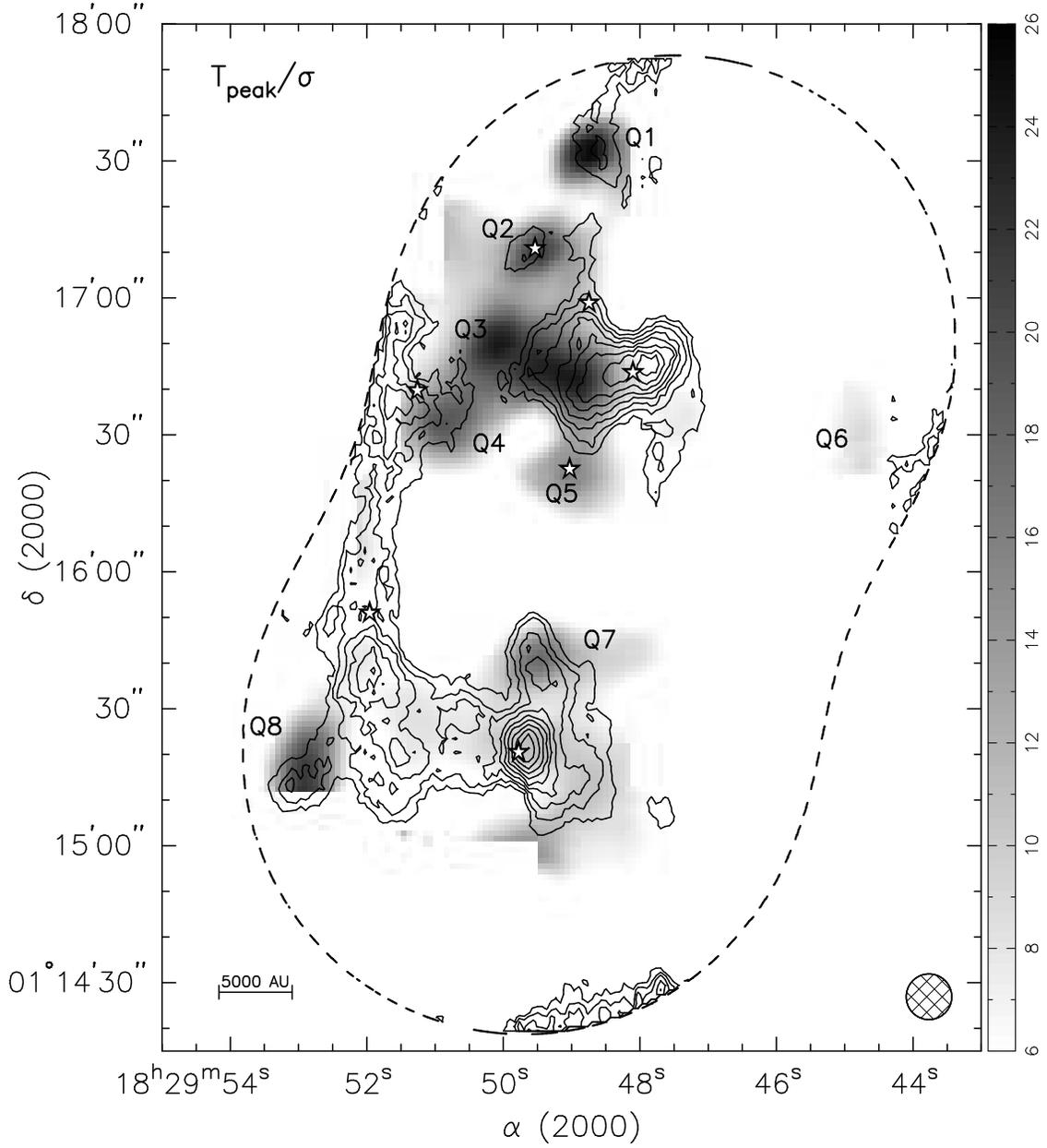}}
\vskip 0.0in
\figcaption[f3.ps]{The peak temperature divided by the velocity dispersion for
the \n2hp\ spectra, a measure of the optical depth, plotted in greyscale
over a contour map of integrated emission.
The greyscale ranges linearly from 6 to 24~K~(km~s\e)\e.
The boundary and annotations are as in Figure~2.
The quiescent cores are most prominent in this map since they possess
both a locally small dispersion and high peak temperature.
\label{fig:peaksig}}
\end{figure}

The eight quiescent cores tend to have higher peak temperatures than
their immediate surroundings and are especially prominent in
Figure~\ref{fig:peaksig} which plots the peak \n2hp\ temperature divided
by the total velocity dispersion. This is a measure of the optical depth
(which was not well determined from the hyperfine fitting) and it rises
toward the quiescent cores which possess both relatively low dispersion
and high peak temperatures.

The quiescent core Q2 is almost coincident with the continuum
source S68Nb and Q5 extends to encompass S68Nd but the
other quiescent cores are apparently starless.
Q6 lies $13''$ southeast of the strong CS core
S68NW that was discussed in Williams \& Myers (1999a).
Table~2 lists the quiescent core locations, non-thermal velocity
dispersion and the ratio of (thermal plus non-thermal) velocity
dispersion as the data are smoothed from $10''$ to $50''$.
This ratio is discussed later in \S\ref{sec:turb} in the context
of pressure gradients.
Spectra toward the core centers at $15''$ and $50''$ resolution
are shown in Figure~\ref{fig:qspec_n2hp}.
The $15''$ spectra are slightly smoothed from the highest available
resolution so as to achieve a higher signal-to-noise ratio and more
clearly show fine features in the spectral profiles.
The lines are narrower and brighter at higher resolution and
resemble the ``kernels'' in cluster forming cores described
in Myers (1998). We discuss this point further in \S\ref{sec:turb}.
The red shoulders in Q1, Q5, and Q8 are also apparent in the other
hyperfine components and may simply be velocity structure or possibly
self-absorption. None of the quiescent core spectra show blue shoulders
from high velocity infall.

The concomitant decrease in velocity dispersion and increase
in optical depth as measured by the peak temperature divided by
the dispersion suggests that the quiescent cores have condensed
out of the larger scale cluster forming cloud through
a localized reduction in turbulent pressure support.
The dissipation of turbulence as a means of core formation
has been alluded to previously but has only recently been
discussed explicitly by Nakano (1998) and Myers \& Lazarian (1998).
A decrease in pressure support should result in an inward flow
of material. The search for such a flow is the subject of the
following subsection.

\subsection{CS(2--1)}
\label{subsec:cs}
A map of CS and \c34s(2--1) spectra from FCRAO observations
is shown in Figure~\ref{fig:fcrao}.
Since these two species share very similar chemical pathways, the
differences in profiles are primarily due to their different optical depths.
The ratio of peak CS to \c34s\ temperatures ranges from of 4--8 indicating
moderate CS optical depths $\tau\simeq 1-3$ (Williams \& Myers 1999b).
Whereas the \c34s\ spectra present a single peak and are approximately
gaussian, the CS spectra have low-level line wing emission and are
generally double-peaked. The line wings are due to outflows from
several cluster members, as we show below, and the two peaks result
from self-absorption since the \c34s\ emission peaks at the dip
of the CS spectra. Self-absorbed spectra from a static core
would be symmetric but here they are clearly asymmetric.
We use the asymmetries in the self-absorption to probe the
velocity differences between outer and inner regions of the cores,
and thereby search for inward motions (e.g. Zhou 1995).
A blue (low velocity) peak that is brighter than the red (high velocity)
indicates that the (outer) absorbing layer is relatively red-shifted,
i.e., infalling, whereas the opposite asymmetry implies outward motions.
The greater the blue-red difference, the greater the relative motions
between the inner and outer regions of the core (Myers et al. 1996).

There is a preponderance of spectra with infall-type asymmetry
and only a few spectra around S68N with the opposite symmetry
in Figure~\ref{fig:fcrao}. Moreover, the average spectrum over the cluster
(not shown) is self-absorbed with a brighter blue than red peak.
This suggests that there is large scale contraction of the gas
around the cluster.
The size of the contracting region is greater than 0.2~pc in diameter
and extends well beyond the continuum sources.
Extended asymmetrical self-absorption in this source in the
$2_{12}-1_{10}$ line of H$_2$CO is also discussed in Mardones (1998).
We have observed a similarly sized infalling region in CS(2--1)
in the Cepheus A cluster forming region (Williams \& Myers 1999b).
At $50''$, the resolution of these data is too coarse to isolate the
dynamics of individual cores but the addition of the interferometer
maps allows us to follow infall and outflow motions down to the scale of
the individual protostellar cores.

The Serpens cloud is known to contain a number of molecular outflows
(White, Casali, \& Eiroa 1995; Davis et al. 1999).
Figure~\ref{fig:outflows} maps the blue- and red-shifted emission
for the BIMA data only.
By analyzing the data prior to combining with the FCRAO data, the bulk of
the mostly uniform cloud emission is resolved out and small scale features,
such as outflows, stand out.  The resulting map shows outflows around
S68N, SMM1, SMM10, and possibly S68Nb. The existence of an outflow
from S68Nb is uncertain because of confusion with the extensive red
lobe around S68N.

\begin{figure}[htpb]
\vskip 0.0in
\centerline{\psfig{figure=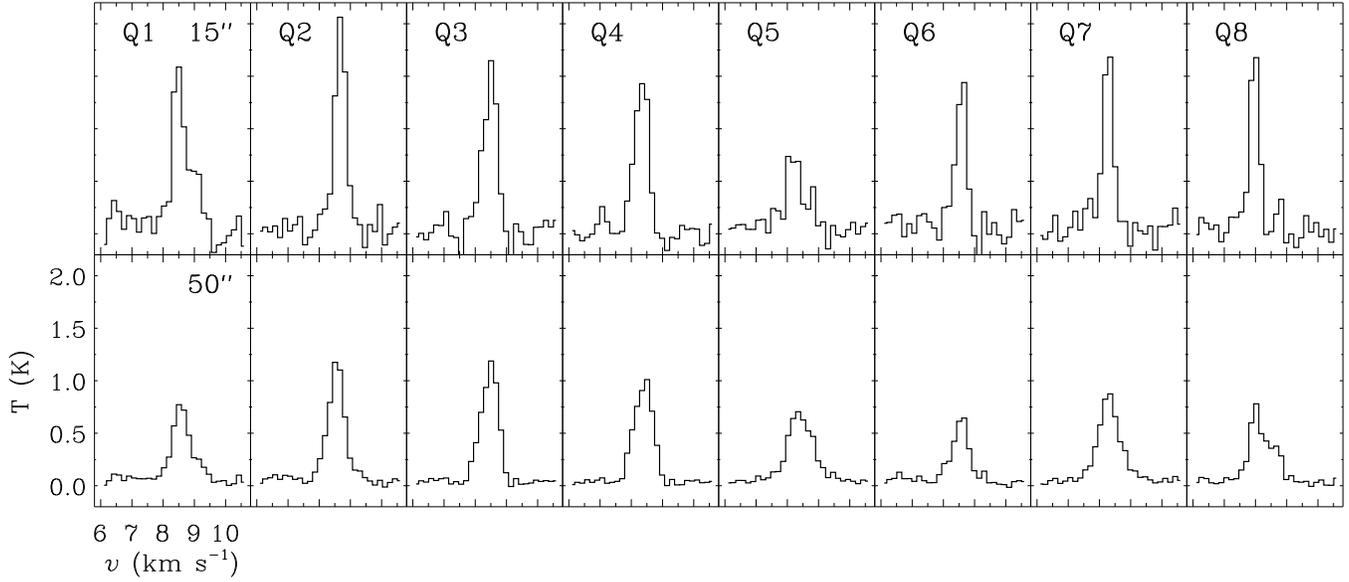,height=5.5in,angle=90,silent=1}}
\vskip -1.8in
\figcaption[f4.ps]{\n2hp\ spectra at the center of the 8 quiescent cores Q1--Q8.
The upper panels show spectra at $15''$ resolution and the
lower panels show spectra at the same positions at $50''$ resolution.
Only the isolated (F$_1$F$=01-12$) hyperfine component is shown.
The velocity and temperature scale is the same for all panels
and is indicated in the lower left.
\label{fig:qspec_n2hp}}
\end{figure}

\begin{figure}[htpb]
\vskip 0.0in
\centerline{\psfig{figure=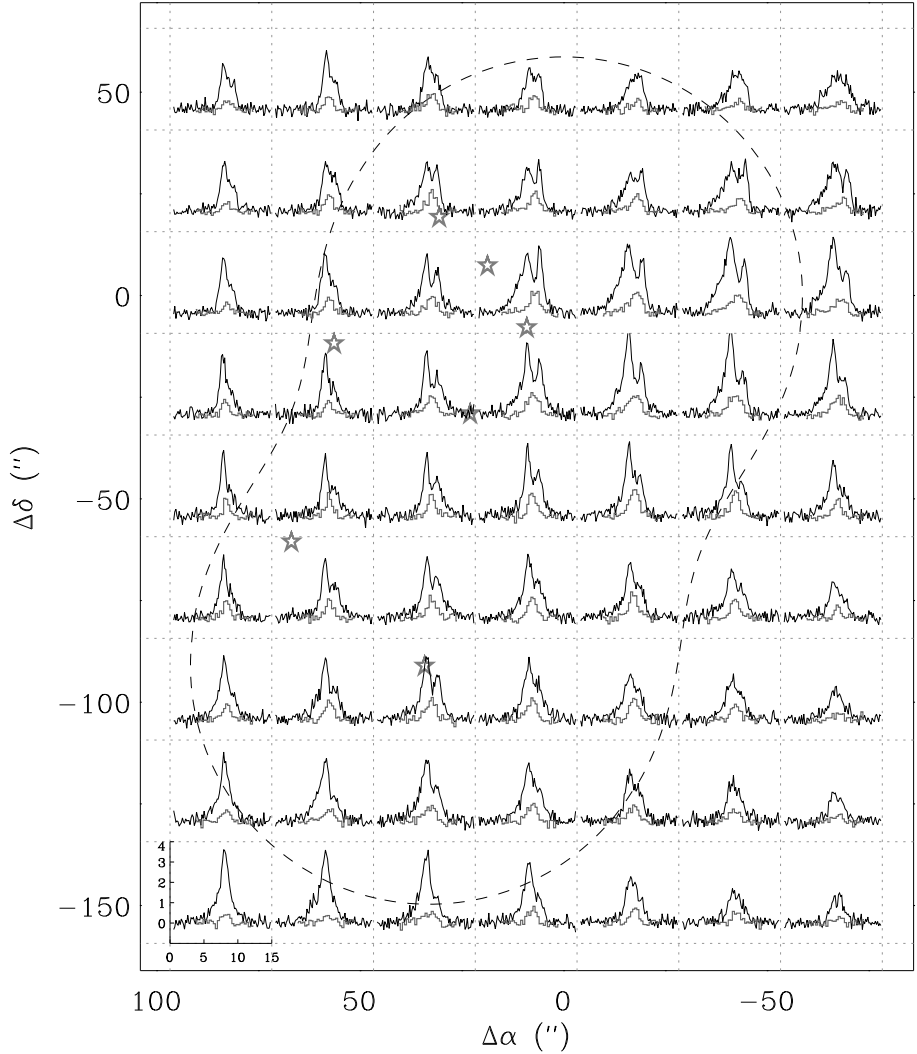,height=6.5in,angle=0,silent=1}}
\vskip 0.0in
\figcaption[f5.ps]{Map of CS and \c34s(2--1) spectra from FCRAO observations
(dark and light lines respectively).
The \c34s\ intensities have been multiplied by a factor of 2 for clarity.
The spectra are placed on a (Nyquist) $25''$ grid with a velocity
range 0 to 15~\kms\ and a temperature range $-1$ to 4~K ($T_R^\ast$)
for each box shown by the dotted line and indicated by the axes for
the lower left box. The outline of the BIMA CS map and the seven
continuum sources from Figure~1 are indicated. Offset coordinates
are relative to $\alpha(2000)=18^{\rm h}29^{\rm m}47\fs 5,
\delta(2000)=01^\circ 16'51\farcs 4$.
\label{fig:fcrao}}
\end{figure}

\begin{figure}[htpb]
\vskip 0.0in
\centerline{\psfig{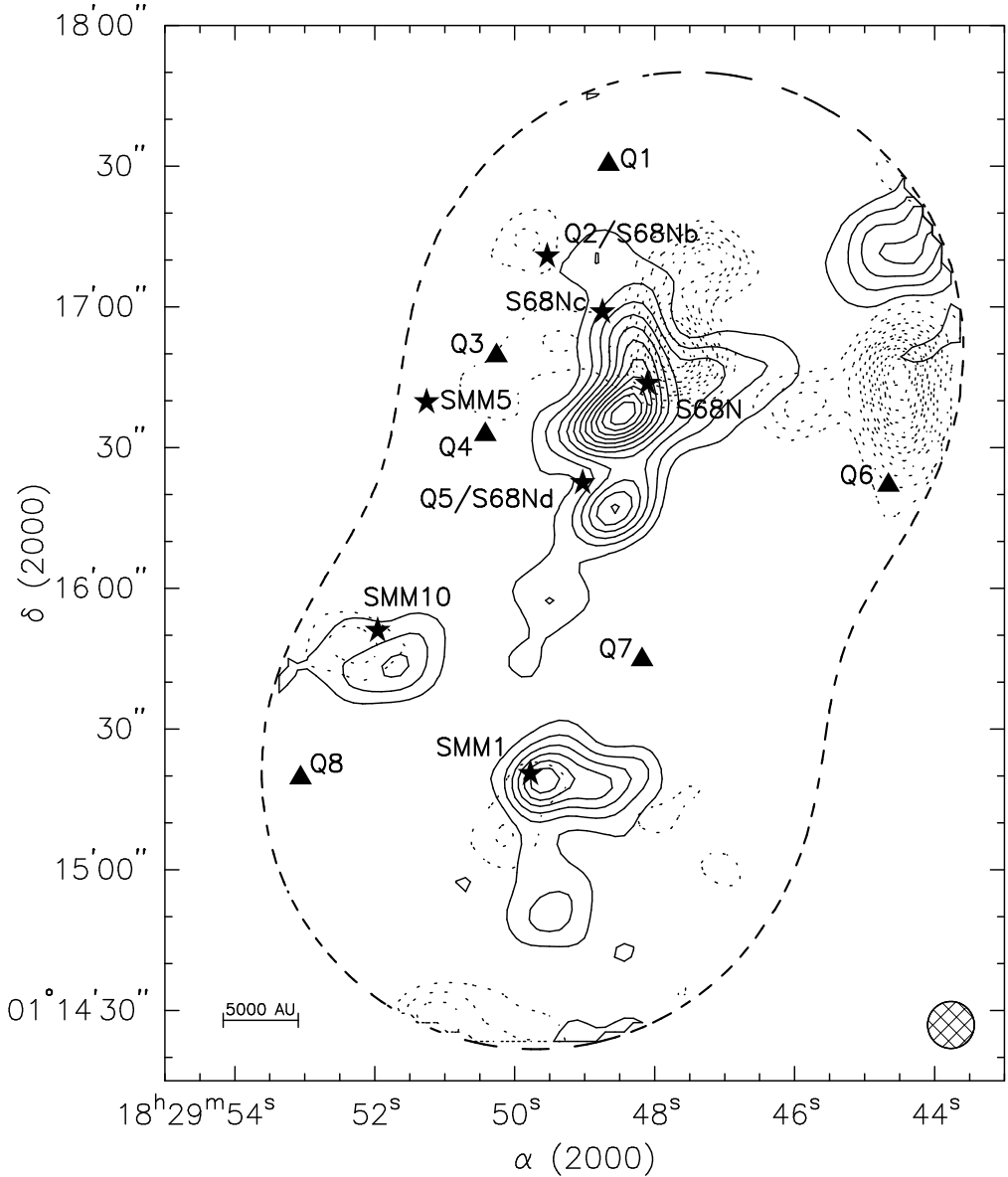}}
\vskip 0.0in
\figcaption[f6.ps]{Blue- and red-shifted CS emission showing several protostellar
outflows. The dotted contours show the integrated intensity from
5.75 to 6.75~\kms, the solid contours from 9.25 to 11.0~\kms.
Starting level and increment is 0.5~K~\kms\ in each case.
The map is made using BIMA data only restored to a circular
$10''$ FWHM beam indicated in the lower right corner.
The locations of the continuum sources and quiescent cores are
indicated by the star and triangle symbols respectively.
The S68N outflow dominates and its red lobe extends across the Q5/S68Nd core.
The SMM1 outflow is confined to a much smaller projected area
and does not appear to be responsible for line wings around other cores.
There is a weak outflow around continuum core SMM10,
and hints of an outflow around Q2/S68Nb.
\label{fig:outflows}}
\end{figure}

\begin{figure}[htpb]
\vskip 0.0in
\centerline{\psfig{figure=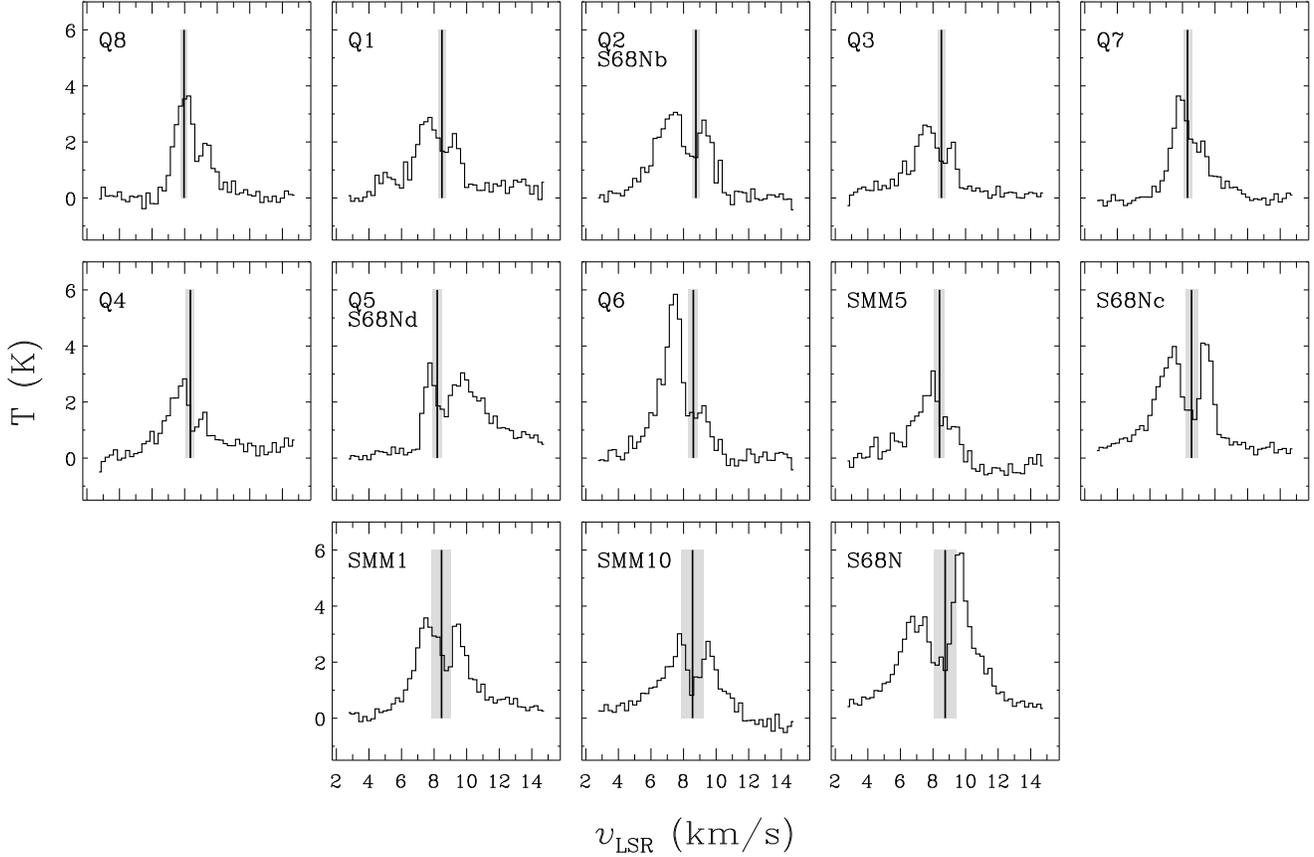,height=5.2in,angle=90,silent=1}}
\vskip 0.0in
\figcaption[f7.ps]{CS spectra toward the quiescent cores and continuum sources.
Spectra have been slightly smoothed to $15''$ to better show the
lineshapes. The velocity and temperature scale is the same for each
spectrum. \n2hp\ velocities and FWHM linewidths are indicated by the
solid vertical line and gray shading respectively; spectra are ordered
from lowest (top left) to highest (bottom right) \n2hp\ linewidth.
\label{fig:qspec_cs}}
\end{figure}

CS spectra toward the quiescent cores and continuum sources are plotted
in Figure~\ref{fig:qspec_cs}. Here, we use the combined FCRAO/BIMA
dataset since it is essential that all there be no missing flux if we
are to interpret the spectra correctly.
Spectra are centered on the position of minimum \n2hp\ linewidth
for the quiescent cores or the continuum source otherwise,
and are at a slightly smoothed $15''$ resolution.
There are approximately symmetric line wings from the
S68N, SMM1, and SMM10 outflows, a blue line wing from the
possible outflow around Q2/S68Nb, and weak one-sided wings
from the red lobe of the S68N outflow around Q5/S68Nd and Q7.
The \n2hp\ velocity and FWHM linewidth are indicated by the solid
vertical line and shading respectively (spectra are shown in
Figure~\ref{fig:qspec_n2hp}).
The \n2hp\ velocity lies at the dip of the double-peaked
CS spectra as for the \c34s\ spectra in Figure~\ref{fig:fcrao}
except for Q8.

The Q8 core, with the narrowest \n2hp\ linewidth in the map,
lies at the edge of the cluster and is uncontaminated by CS outflow
wings. The CS spectra around the core all have the classic
infall profile, but at the position of the linewidth minimum
the \n2hp\ velocity lines up with the blue CS peak and not the dip
as would be expected for self-absorption. Given that all the other
double-peaked spectra in the map are self-absorbed it seems unlikely
that the double-peaked spectra in this region are not also
self-absorbed. However, the central \n2hp\ spectrum in
Figure~\ref{fig:qspec_n2hp} shows a second, red, peak at $15''$
and a red shoulder at $50''$ and it may be that the \n2hp\ is also
self-absorbed here.
The low resolution \c34s\ data in Figure~\ref{fig:fcrao}
demonstrates that the CS spectra are self-absorbed in this region but
to settle the issue in the Q8 core itself, sensitive higher resolution
observations of \c34s\ are necessary.

Aside from the uncertainty over the interpretation of the Q8 core,
Figure~\ref{fig:qspec_cs} shows an overall trend for the
CS infall asymmetry to be greatest in those cores
with small \n2hp\ linewidths. That is, the CS spectra toward
the quiescent cores all have a greater blue
than red peak indicating a positive infall velocity,
but cores S68Nc, SMM10, and SMM1 present a very symmetric
appearance indicating near-zero infall.
The extended red CS outflow lobe from S68N in Figure~\ref{fig:outflows}
makes it difficult to isolate the Q5/S68Nd and Q7 cores and
study their dynamics individually in this line, even though they
are well separated in the \n2hp\ map.
It is particularly difficult to diagnose infall motions around
S68N itself where the infall asymmetry abruptly reverses from one
position to another $10''$ away. Nevertheless, Figure~\ref{fig:qspec_cs}
suggests a connection between the level of core turbulence and
the asymmetry of the CS profiles, in turn related to the infall speed.
We explore this in more detail in the following section.

\section{Pressure driven flows and turbulent dissipation}
\label{sec:turb}
The singledish spectra in Figure~\ref{fig:fcrao} suggests
a global infall onto the cluster but the addition
of the interferometer data allows a higher resolution view that
reveal a wide range of spectral asymmetries
that present a patchwork of inward and outward flows.
The size of the region over which infall motions are observed
is large, $\sim 0.2$~pc in diameter, at least partly because
the cluster contains a number of collapsing cores.
The ($\sim 3000$~AU) resolution of the combined dataset separates
out the individual protostars and protostellar cores from each other,
and permits an analysis of how the small scale inward and outward
motions are related to the local environment.

Detailed modeling of line profiles enable the infall speed to
be determined (e.g., Zhou 1995; Williams et al. 1999)
in isolated, low mass star forming cores
but there is a much greater range of spectral shapes in the CS data here.
For example, many are contaminated by outflows, either internal or
overlapping from neighboring sources. Thus the modeling requires
several additional parameters with the result that the uncertainty in
the determination of the infall speed at each point is large.
Therefore, rather than try to fit the spectra directly,
we estimated the location of the main features
in the CS spectra by eye using a simple cursor based routine.
This resulted in a catalog of positions and velocities where the CS
spectra have local peaks and dips.

The simplest estimate of the CS spectral asymmetry is the difference in
blue and red peak temperatures, $\dtbr=T_{\rm blue}-T_{\rm red}$.
We consider the difference, rather than the
ratio of peak temperatures since it is determined with a smaller error.
This is obviously only defined for double-peaked spectra and we therefore
exclude spectra with possible infall ``shoulders''.
A second measure of asymmetry that has the advantage of being defined
for all spectral shapes is the velocity difference between the peak
CS and \n2hp\ velocities. This is the unnormalized
$\delta v$ parameter introduced by Mardones et al. (1997).
Based on simulations with simple two-layer infall models (Myers et al. 1996),
we find that the blue-red temperature difference correlates linearly
with the infall speed for a given optical depth and excitation temperature
and that the scatter when a range of optical depths $\tau=2-5$
and excitation temperatures $T_{\rm ex}=15-25$~K
is considered is relatively small.
The velocity difference also correlates linearly with infall speed
but is more sensitive to optical depth variations.
Consequently we use \dtbr\ as a measure of infall
speed and analyze its distribution across the cluster.

\begin{figure}[htpb]
\vskip 0.0in
\centerline{\psfig{figure=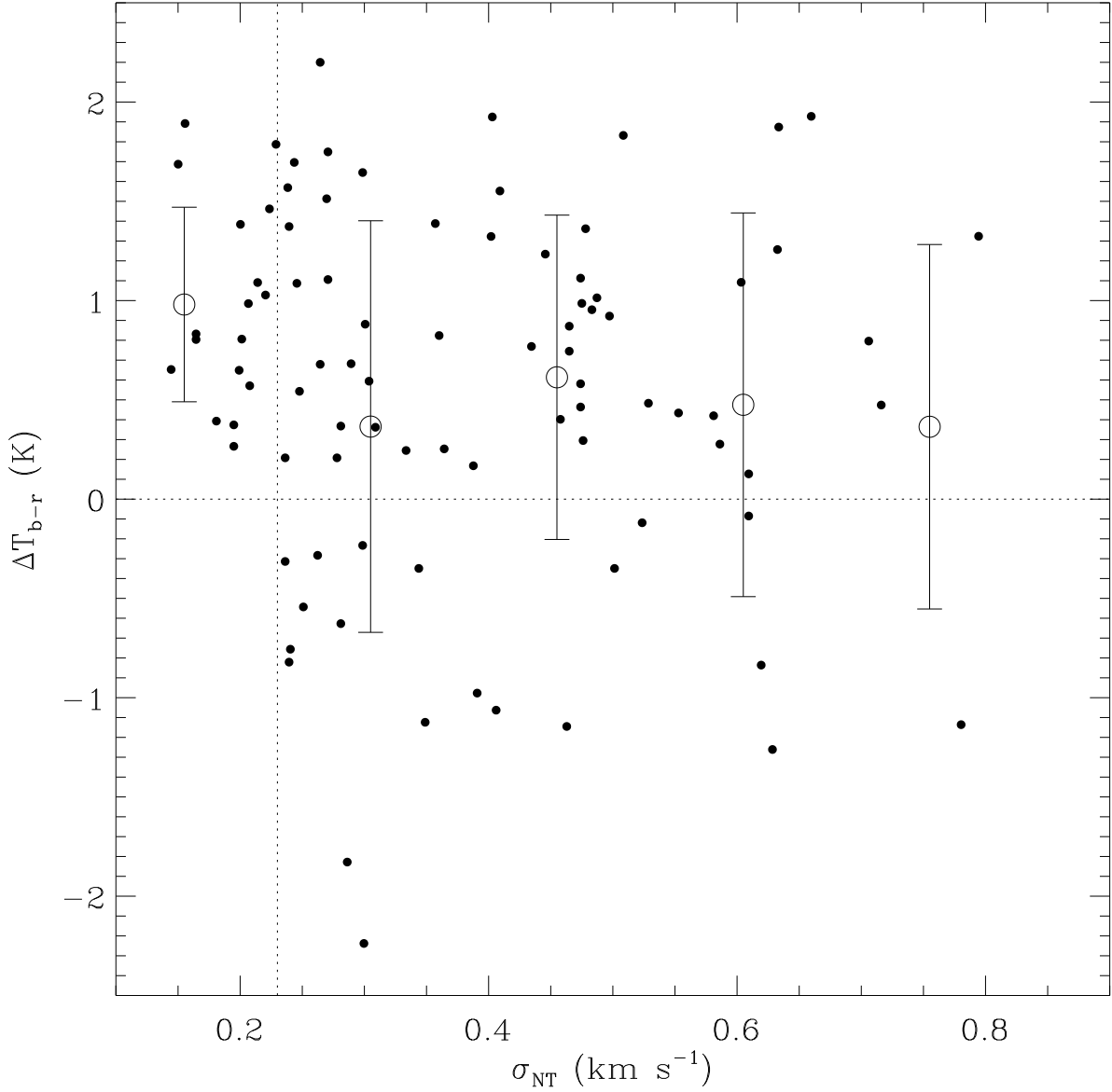,height=6.8in,angle=0,silent=1}}
\vskip -0.1in
\figcaption[f8.ps]{The CS blue-red temperature difference plotted against the
\n2hp\ non-thermal velocity dispersion
for all points in the dataset where both the CS spectra possess two peaks
and the \n2hp\ spectra have a peak signal-to-noise ratio greater than 5.
The vertical dotted line is drawn at $\sigma_{\rm NT}=0.23$~\kms. All
spectra with lower values of dispersion have positive values of \dtbr\
indicating a positive infall speed. The large open circles and error
bars indicate the mean and standard deviation of the temperature
difference for five velocity dispersion bins.
\label{fig:deltat}}
\end{figure}

To test the hypothesis made in the previous section
that the CS asymmetries are large where
the \n2hp\ velocity dispersion is small, we simply plot \dtbr\
against $\sigma_{\rm NT}(\n2hp)$ in Figure~\ref{fig:deltat}.
The points are very scattered and there is no significant correlation
between the two quantities. However, when binned by $\sigma_{\rm NT}$, the
average \dtbr\ is greater than zero (implying a positive infall velocity).
Moreover, $\dtbr >0$~K for {\it all} spectra with $\sigma_{\rm NT}<0.23$~\kms.
The lack of direct correlation is because several different dynamical
states have been grouped together. By selecting individual regions,
we can isolate the different states.

\begin{figure}[htpb]
\vskip 0.0in
\centerline{\psfig{figure=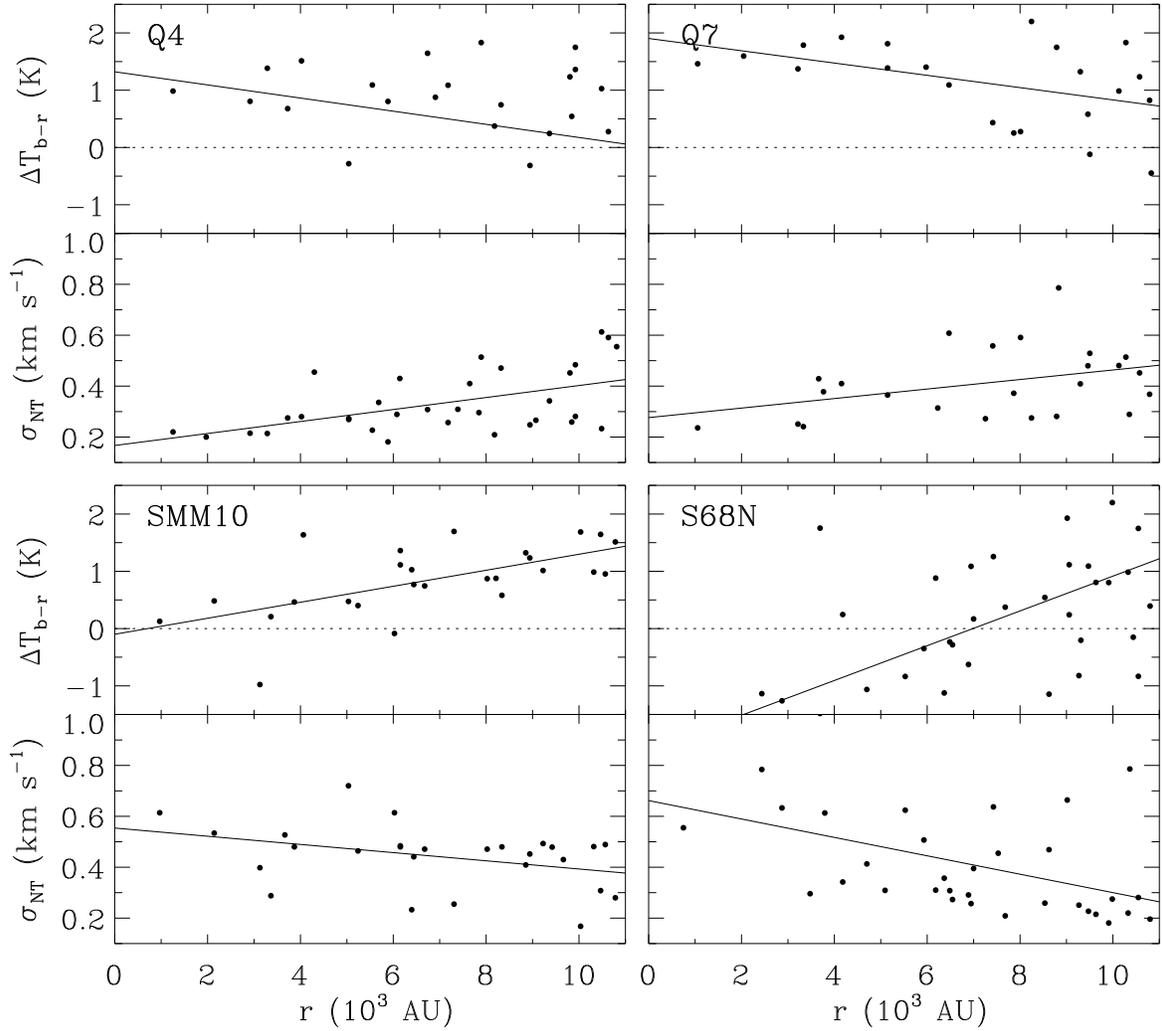,height=6.0in,angle=90,silent=1}}
\vskip -0.1in
\figcaption[f9.ps]{The radial variation of the CS blue-red temperature difference
and \n2hp\ non-thermal velocity dispersion for two quiescent cores
and two continuum sources. Points from individual spectra and a least
squares fit are shown.
\label{fig:radial}}
\end{figure}

Figure~\ref{fig:radial} plots the variation of CS asymmetry and \n2hp\
velocity dispersion against the distance from the center of a core
for two quiescent cores and two continuum sources.
A similar behavior is found in the other cores:
the velocity dispersion tends to increase and the blue-red
temperature difference generally decreases with increasing
radius from the center of the quiescent cores and vice versa
for the continuum sources.
Moreover, the temperature difference is positive at the
centers of the quiescent cores but is small or negative
(i.e., outflow) at the centers of the continuum cores.

\begin{figure}[htpb]
\vskip 0.0in
\centerline{\psfig{figure=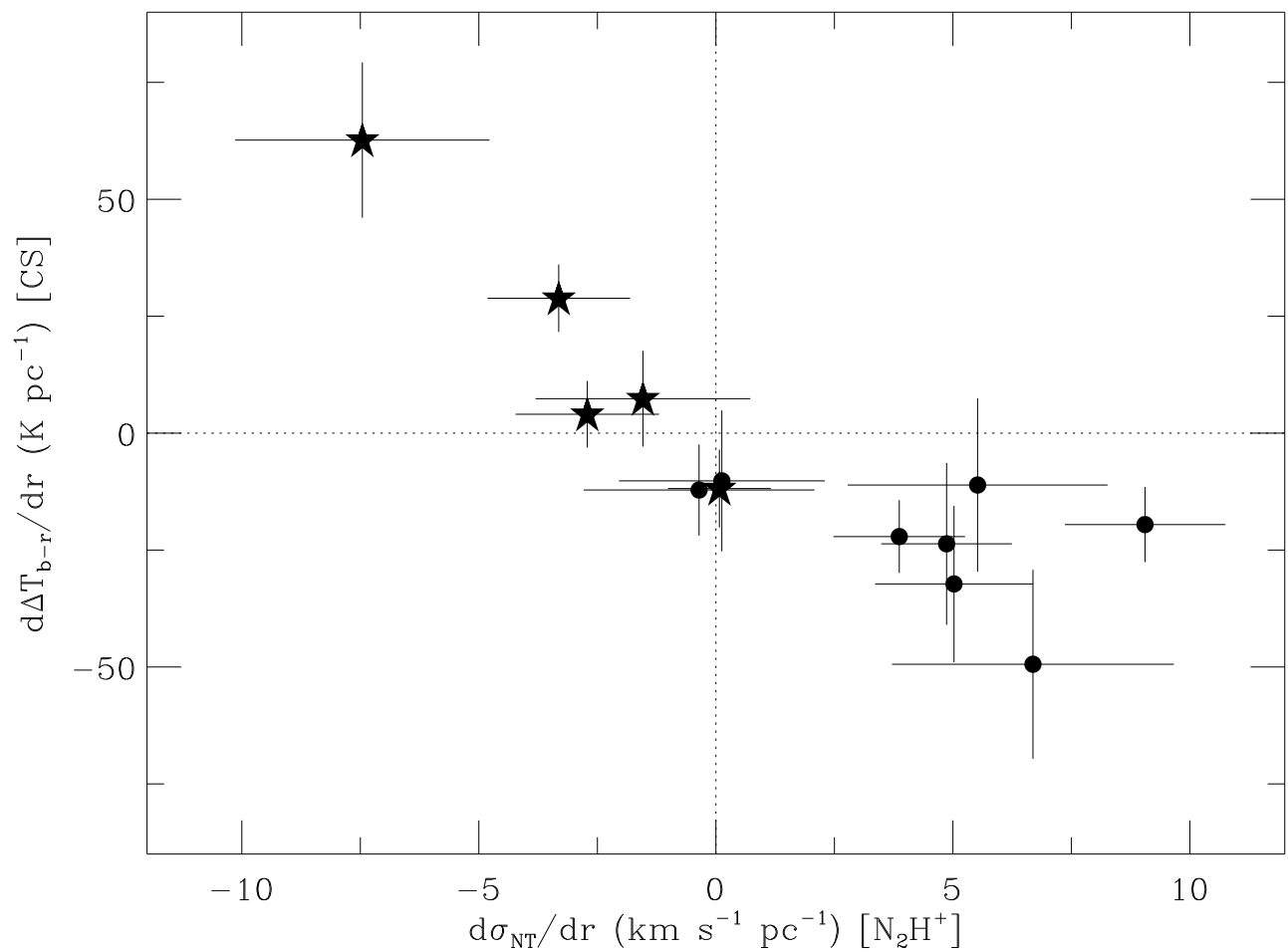,height=5.4in,angle=90,silent=1}}
\vskip -0.1in
\figcaption[f10.ps]{The radial gradient of \dtbr(CS) plotted against the radial
gradient of $\sigma_{\rm NT}$(\n2hp) for the continuum sources (stars)
and quiescent cores (circles). The gradients are the slopes of
least squares fits as in Figure~\ref{fig:radial}.
\label{fig:radgrad}}
\end{figure}

The slopes of the least squares fits, or the radial gradients, of the
CS blue-red temperature difference and \n2hp\ non-thermal velocity
dispersion are tabulated in Table~3
and plotted against each other in Figure~\ref{fig:radgrad}.
Generally, the quiescent cores lie toward the lower right section
(increasing \n2hp\ dispersion and decreasing CS asymmetry with radius)
and the continuum sources lie in the upper right section
(decreasing \n2hp\ dispersion and increasing CS asymmetry with radius).
For a constant density, a change in velocity dispersion implies
a change in pressure which results in a flow toward the lower
pressure (i.e., lower dispersion) regions.
The conversion from CS asymmetry to infall speed
depends not only on the blue-red temperature difference but also
the excitation temperature and optical depth. If these do not vary
greatly from core envelope to center then the observed increase in
\dtbr\ as $\sigma_{\rm NT}$ decreases toward the centers of the
quiescent cores implies an increase in infall speed.

Since the pressure depends linearly on the density and quadratically
on the velocity dispersion, and since the infall speed is more
sensitive to the blue-red temperature difference than the excitation
temperature and optical depth, the inverse correlation of \dtbr\
with $\sigma_{\rm NT}$ in Figure~\ref{fig:radgrad}
is evidence for pressure driven, inward, flows in the quiescent cores.
The correlation also extends to negative dispersion gradients and outflow
motions around the continuum sources illustrating the disruptive effect
of young protostars on their parent cloud.
Finally, we note that a fit through the
data points suggests a slightly negative CS asymmetry gradient
at $d\sigma_{\rm NT}/dr=0$ which may indicate small, presumably
gravitational, inward motions even when the pressure gradient is zero.

Using the two-layer model of Myers et al. (1996), we find that
the infall speed corresponding to a typical value, $\dtbr =1$~K,
for $\sigma_{\rm NT}=0.3$~\kms\ is $v_{\rm in}\simeq 0.05$~\kms.
This is quite small and similar to that expected for the quasistatic
contraction of a magnetically supported, isothermal core (Lizano \& Shu 1989).
However, the nature of the flow is very different 
from the predictions of such ambipolar diffusion models:
all the cores, whether with or without continuum sources,
have highly non-thermal linewidths and the infall speed is
greatest in those cores with the smallest linewidths.
This inverse correlation is not expected in a purely gravitational
collapse and is more suggestive of a pressure driven flow.
Nakano (1998) and Myers \& Lazarian (1998) describe how
inward flows onto a core can occur through the dissipation
of turbulence and consequent loss of pressure support.
As a core grows and its center becomes more opaque to ionizing
radiation, its coupling to the magnetic field, and the range of
MHD waves that propagate through it, decrease.
Without replenishment, the waves decay within a free-fall time
and the turbulent pressure, $\rho\sigma_{\rm NT}^2$, rapidly decreases
(where $\rho$ is the mass density).
Since wave support is maintained in the lower opacity,
more highly ionized, core envelope, its pressure remains the same
with the result that there is a pressure gradient, leading to a flow,
from core envelope to center. The magnitude of the flow
will be greatest where the pressure gradient is greatest,
or equivalently where the central linewidths are smallest
if the external pressure is approximately constant.

The magnitude of the inward speed depends on the pressure
difference between core center and envelope.
The observed inward motions are small, $v_{\rm in}\ll\sigma_{\rm NT}$,
and therefore the pressure differences are small.
Since the total (thermal plus non-thermal) velocity dispersion
at $10''$ is a factor of $0.36-0.75$ less than at $50''$ (Table 2),
the density contrast between core centers and envelopes
is inferred to lie in the range $2-8$.
This is comparable to the ratio of peak to average density for cores
in the Ophiuchus cluster forming cloud (Motte, Andr\'{e}, \& Neri 1998).

As the core grows, the opacity increases further until the
ionization is dominated by cosmic rays ($A_V\simeq 4$; McKee 1989).
In this case, Myers (1998) predicts the existence of ``kernels'',
$\sim 6000$~AU, in size that are completely cutoff from MHD waves.
If the external pressure is sufficiently large, as in massive
star forming regions, these kernels can be stable, supported
by thermal pressure against their self-gravity.
The criterion for stability (Myers 1998 equation 3)
can be rewritten as $\sigma_{\rm NT}/\sigma_{\rm T}=1.5$
which is satisfied in all the cores here where we have found
$\sigma_{\rm NT}/\sigma_{\rm T}>2.1$.
The quiescent cores extend over $10''-20''$, which is close
to the expected size of a kernel, and their velocity FWHM
are $\sim 0.5$~\kms\ similar to Myers' Figure~2, but there is
insufficient signal-to-noise to discern the predicted thermal ``spike''.
There are possible examples in the residuals
to the hyperfine fits to the spectra but they are not consistent
across all the components and may be due to poorly cleaned sidelobes
from other cores (the cleaning method is non-linear and varies
in its effectiveness from channel to channel).
High spatial and velocity resolution singledish observations of
higher transition lines such as \n2hp(3--2) offer an independent
test for the presence of a thermal spike and may also be used to
constrain the density contrast in the cores.

The maps in Figures~\ref{fig:bima},\ref{fig:sig_nt}
reveal a number of protostars and pre-protostellar collapsing cores.
The cluster did not form in a single event, therefore, but continues to
accrue members through an ongoing process of individual star formation.
Hurt \& Barsony (1996) analyzed the spectral energy distribution (SED)
of several of the bright sources in this cluster and concluded that they
were Class 0 protostars. The IRAS data does not have the resolution to
resolve the emission (and therefore to define their SED in the far-infrared)
from the seven continuum sources that we have identified here
but if, following Hurt \& Barsony, we divide up the IRAS fluxes
evenly between all the objects, all seven would be classified as Class 0.

The discovery of the quiescent cores, and their association with
high infall motions, suggests that they are the precursors to the 
Class 0 sources. Within the boundaries of the maps here, and at the
sensitivity of the observations, we have found approximately equal
numbers of continuum sources and quiescent cores (7 and 8 respectively,
with 2 shared). If stars continue to form at a constant rate, then
the lifetime of the quiescent cores must be approximately the same
as the lifetime of the Class 0 phase of protostellar evolution,
$\sim 3\times 10^4$~yr (Andr\'{e} \& Montmerle 1994).
Such a short lifetime implies a dynamic evolution since
the free-fall timescale,
$t_{\rm ff}=(G\rho)^{-1/2}\simeq 6\times 10^4$~yr,
for $\nh2=10^6$~\cc, approximately equal to the inferred
volume density of the quiescent cores and a factor of two greater
than the critical density of \n2hp.
This rapid evolution is consistent with core growth through the decay
of turbulence since this should occur on a free-fall timescale (Nakano 1998).

\section{Summary}
This paper presents millimeter wavelength continuum and spectral line
observations of a young, embedded, low mass cluster forming region
in the Serpens molecular cloud. 7 continuum sources are found at the 
resolution and sensitivity of these data. The distribution of these
sources corresponds well with the \n2hp\ emission but
the CS data presents a different appearance, with high velocity
emission from outflows around 4 continuum sources, and central
dips in spectral profiles from self-absorption (as shown on the
large scale from singledish \c34s\ data and on the small scale
from the \n2hp). Away from the powerful outflow around S68N, the
self-absorption is red-shifted which we interpret as indicating
inward motions.

The \n2hp\ linewidth is dominated by non-thermal motions throughout
the cluster. However, there are 8 regions where the non-thermal velocity
dispersion reaches a local minimum. 6 are starless and 2 contain continuum
sources.  They do not all coincide with peaks of the integrated intensity
but they all stand out in maps of the peak temperature divided by the
dispersion, a measure of the optical depth.
The CS spectra toward these ``quiescent'' cores are particularly
asymmetric, indicating relatively high infall speeds.
Generally, the \n2hp\ dispersion increases and the CS blue-red temperature
difference decreases with increasing distance from the core centers, and
vice versa for the continuum sources. The correlation of CS asymmetry,
related to infall speed, with \n2hp\ dispersion, related to the local
turbulent pressure, suggests that the inward flows are at least partly
pressure driven and that the cores formed through the localized dissipation
of turbulence as envisioned by Nakano (1998) and Myers \& Lazarian (1998).
Such a scenario is consistent with the observed numbers of quiescent
cores and Class 0 sources.

The singledish data alone shows a net inward motion onto the cluster.
Although there is clearly considerable smearing of the detailed dynamics,
this suggests that it may be fruitful to search for infall signatures
in more distant clusters at $\sim 0.1-0.2$~pc resolution.
It will, of course, be important to observe other nearby cluster
forming regions at higher resolution, $\simlt 3000$~AU, to augment
this study. Ultimately, the comparison of conditions in many
clusters will give a clearer picture of their formation,
show the effects of different environments, and, through an inventory
of continuum sources and pre-protostellar cores, can be hoped to elucidate
the origins of the stellar IMF (Motte et al. 1998; Testi \& Sargent 1998).

\acknowledgments
JPW is supported by a Jansky fellowship. Partial support has also been
provided by the NASA Origins of Solar Systems Program, grant NAGW-3401.
We thank Leo Blitz and Dick Plambeck for generously assigning additional
BIMA tracks to remap the continuum emission and Marc Pound and
Tamara Helfer for advice concerning the data reduction.

%\clearpage
% ------------------------ REFERENCES --------------------

\clearpage
% ------------------------ END REFERENCES --------------------

% ------------------------ TABLES --------------------
\begin{table}
\begin{center}
TABLE 1\\
Continuum Sources\\
\vskip 2mm
\begin{tabular}{lrrrr}
\hline\\[-2mm]
Source & $\Delta\alpha^{\rm a}$ & $\Delta\delta^{\rm a}$
       & Flux density & Peak \\
       & ($''$) & ($''$) & (mJy) & (mJy~beam$^{-1}$)\\[1mm]
\hline\hline\\[-3mm]
SMM1            & 34.1 & --91.0 & 233.7 & 164.7 \\
S68N            &  9.0 &  --7.8 &  36.1 &  20.2 \\
S68Nd           & 23.0 & --28.9 &  32.3 &  15.6 \\
SMM10           & 66.9 & --60.4 &  25.3 &  22.1 \\
SMM5            & 56.4 & --11.7 &  20.7 &  13.7 \\
S68Nc$^{\rm b}$ & 18.7 &    7.5 &  20.4 &  12.1 \\
S68Nb           & 30.6 &   19.4 &  17.7 &  15.1 \\[2mm]
\hline\\[-2mm]
\multicolumn{5}{l}{$^{\rm a}$ location of peak relative to}\\
\multicolumn{5}{l}{$^{\rm ~}$
$\alpha(2000)=18^{\rm h}29^{\rm m}47\fs 5,
\delta(2000)=01^\circ 16'51\farcs 4$}\\
\multicolumn{5}{l}{$^{\rm b}$ possibly multiple sources}\\
\end{tabular}
\end{center}
\label{tab:continuum}
\end{table}

\begin{table}
\begin{center}
TABLE 2\\
Quiescent Cores\\
\vskip 2mm
\begin{tabular}{lrrcccc}
\hline\\[-2mm]
Core   & $\Delta\alpha^{\rm a}$ & $\Delta\delta^{\rm a}$
       & $v_{\rm LSR}$ & $\sigma_{\rm NT}$ ($10''$)
       & ${\sigma_{\rm tot}(10'')\over\sigma_{\rm tot}(50'')}$
       & ${T_{\rm peak}(10'')\over T_{\rm peak}(50'')}$ \\
       & ($''$) & ($''$) & (km s$^{-1}$) & (km s$^{-1}$) & & \\[1mm]
\hline\hline\\[-3mm]
 Q1    &   17.4  &   39.0  &  8.482  &  0.17  &  0.52  & 2.9 \\
 Q2    &   28.2  &   18.6  &  8.757  &  0.20  &  0.68  & 1.7 \\
 Q3    &   41.4  &  --1.8  &  8.514  &  0.19  &  0.68  & 1.6 \\
 Q4    &   43.8  & --18.6  &  8.353  &  0.21  &  0.69  & 1.4 \\
 Q5    &   15.0  & --35.4  &  8.189  &  0.23  &  0.52  & 1.2 \\
 Q6    & --42.6  & --29.4  &  8.601  &  0.24  &  0.75  & 1.6 \\
 Q7    &   10.2  & --66.6  &  8.311  &  0.20  &  0.52  & 1.1 \\
 Q8    &   83.4  &  -91.8  &  7.964  &  0.16  &  0.36  & 1.9 \\[2mm]
\hline\\[-2mm]
\multicolumn{7}{l}{$^{\rm a}$
location of minimum $\sigma_{\rm NT}$ relative to}\\
\multicolumn{7}{l}{$^{\rm ~}$
$\alpha(2000)=18^{\rm h}29^{\rm m}47\fs 5,
\delta(2000)=01^\circ 16'51\farcs 4$}\\
\end{tabular}
\end{center}
\label{tab:quiescent}
\end{table}
\clearpage

\begin{table}
\begin{center}
TABLE 3\\
Radial Gradients\\
\vskip 2mm
\begin{tabular}{lrrrr}
\hline\\[-2mm]
Core   & $d\Delta T_{\rm b-r}/dr$ & err
       & $d\sigma_{\rm NT}/dr$ & err \\
       & \multicolumn{2}{c}{(K pc$^{-1}$)}
       & \multicolumn{2}{c}{(km s$^{-1}$ pc$^{-1}$)} \\[1mm]
\hline\hline\\[-3mm]
Q1        &  --12.2  &   9.7  &  --0.36  &  2.43  \\
Q2/S68Nb  &  --11.1  &  18.5  &    5.52  &  2.74  \\
Q3        &  --32.2  &  16.7  &    5.02  &  1.67  \\
Q4        &  --23.7  &  17.3  &    4.87  &  1.38  \\
Q5/S68Nd  &  --10.2  &  15.1  &    0.13  &  2.17  \\
Q6        &  --49.4  &  20.2  &    6.69  &  2.97  \\
Q7        &  --22.1  &   7.8  &    3.87  &  1.39  \\
Q8        &  --19.5  &   8.0  &    9.05  &  1.69  \\
SMM1      &     4.0  &   7.1  &  --2.71  &  1.51  \\
S68N      &    62.7  &  16.6  &  --7.46  &  2.68  \\
SMM10     &    28.8  &   7.2  &  --3.32  &  1.51  \\
SMM5      &  --11.9  &   8.3  &    0.07  &  1.09  \\
S68Nc     &     7.4  &  10.2  &  --1.54  &  2.27  \\[2mm]
\hline\\[-2mm]
\end{tabular}
\end{center}
\label{tab:radial}
\end{table}
\clearpage

% ------------------------ FIGURES --------------------
\clearpage

\clearpage

\clearpage

\clearpage

\clearpage

\clearpage

\clearpage

\clearpage

\clearpage

\clearpage

\end{document}